\documentclass{article}
\usepackage{spconf,amsmath,graphicx}
\usepackage{amssymb}
\usepackage{enumerate}
\usepackage{color}
\usepackage{booktabs}
\usepackage{multirow}
\usepackage{mathtools}
\usepackage{float} 

\title{Graph Spectral Characterization of Brain Cortical Morphology}
\name{Sevil~Maghsadhagh$^{1}$, 
Anders Eklund$^{2,3}$, Hamid~Behjat$^{*4}$ 
\thanks{* corresponding author; e-mail: hamid.behjat@bme.lth.se}
} 

\address{
\small $^{1}$ Independent researcher, Muthgasee 66, 1190 Vienna, Austria 
\\ 
\small $^{2}$ Department of Biomedical Engineering, Link\"{o}ping University, Link\"{o}ping, Sweden
\\
\small $^{3}$ Department of Computer and Information Science, Link\"{o}ping University, Link\"{o}ping, Sweden
\\
\small $^{4}$ Department of Biomedical Engineering, Lund University, Lund, Sweden}

\begin{document}
%
\maketitle
\begin{abstract}
The human brain cortical layer has a convoluted morphology that is unique to each individual. Characterization of the cortical morphology is necessary in longitudinal studies of structural brain change,  as well as in discriminating individuals in health and disease. A method for encoding the cortical morphology in the form of a graph is presented. The design of graphs that encode the global cerebral hemisphere cortices as well as localized cortical regions is proposed. Spectral metrics derived from these graphs are then studied and proposed as descriptors of cortical morphology. As proof-of-concept of their applicability in characterizing cortical morphology, the metrics are studied in the context of hemispheric asymmetry as well as gender dependent discrimination of cortical morphology.
\end{abstract}
\section{Introduction} 
\label{sec:intro}
The conventional approach for characterization of brain morphology and study of its changes is to quantify the volumes of a set of brain structures \cite{Decarli2005}. Cortical thickness measures are also popular means for characterizing morphology \cite{Fischl2000}. As volume and cortical thickness measures are incapable of capturing the full anatomical information, anatomical shape descriptors \cite{Gerardin2009, Shen2012} have been proposed to provide significant complementary informative representation of brain morphology. For example in \cite{Wachinger2015}, it was shown that shape descriptors of cortical and an ensemble of subcortical structures provide a powerful means to discriminate individuals based on their age, sex and neurodegenerative disorder. These shape descriptors use triangular surface mesh or tetrahedral volume tessellation constructions of brain structures, and exploit eigenfunctions of the LaplaceÐBeltrami operator \cite{Reuter2006}. 

Here we build on these works in several respects. Firstly, we use voxel-based graph designs. That is, we use the volumetric voxel representation of the 3D structure of the cortical ribbon, and construct a graph with vertices associated to individual voxels, and connectivities defined based on geodesic adjacencies;  designs of graphs based on a similar encoding of gray matter include, subject-specific designs of cerebral \cite{Behjat2013} and cerebellar \cite{Behjat2014} cortices and group-based template designs \cite{Behjat2015}, but these graphs were leveraged for analysis of fMRI data rather than shape characterization. Secondly, we encode and exploit morphological information of an ensemble of localized cortical regions as opposed to using shape descriptors of the global cortical structure. This is done by designing graphs that encode localized cortical regions. Thirdly, we propose the use of shape descriptors across different spectral bands in contrast to using exact graph Laplacian eigenvalues.  

\section{Methods}
\label{sec:methods}

\subsection{Graphs and Their Spectra}
An undirected, unweighted graph $\mathcal{G}=(\mathcal{V},\mathcal{E},A)$ consists of a set $\mathcal{V}$ of $N_{g}:=|\mathcal{V}|$ vertices, a set $\mathcal{E}$ of edges  (i.e., pairs ($i,j$) where $i,j \in \mathcal{V}$), which can be fully described by an adajcency matrix $A$ with elements $A_{i,j}$ equal to $1$ if  $\text{if} \: (i,j) \in \mathcal{E}$, and $0$, if otherwise.

Using $A$, the graph's diagonal, degree matrix $D$ is defined with elements $ D_{i,i}=\sum_{j} A_{i,j}$, and the graph's normalized Laplacian matrix $L$ is defined as 
\begin{align} 
L & =  I - D^{-1/2} A D^{-1/2}.
\label{eq:L}
\end{align} 

\noindent
Since $L$ is symmetric and positive semi-definite, it can be diagonalized as $
L =  \Sigma \Lambda {\Sigma}^{T},
$ where $ 
\Sigma = [\chi_{_1} | \chi_{_2} | \cdots | \chi_{_{N_{g}}}],
$ is an orthonormal matrix containing a set of $N_g$ eigenvectors $\{\chi_{_k}\}_{k=1}^{N_{g}}$, and $\Lambda$ is a diagonal matrix whose entries equal the associated real, non-negative eigenvalues that define the graph spectrum $\mathcal{S}$ as
\begin{equation}
\label{eq:graphSpectrum}
\mathcal{S} = \text{diag}(\Lambda) = \{ 0 = \lambda_{1} \le \lambda_{2} \le \cdots \le  \lambda_{N_{g}} \le 2 \}.
\end{equation}
Unlike classical Euclidean domain spectrum, each graph has a unique definition of spectrum, with a unique range $[0,  \lambda_{N_{g}}]$ and a unique set of irregularly spaced eigenvalues with possibility of multiplicity greater than one. 

\subsection{Cerebral Hemisphere Cortex Graphs}
\subsubsection{Global Cerebral Hemisphere Cortex (GCHC) Graphs}
For a given hemisphere, a graph that encodes its cortical topology is designed. Cortical ribbons 
extracted using the FreeSurfer software package \cite{Fischl2012} serve as the base of the design. Voxels within the cortical ribbon are treated as graph vertices. Graph edges are defined based on 26-neighborhood connectivity of voxels in 3D space. Two vertices are connected through an edge if they lie within each-other's 26-neighborhood. Due to limited voxel resolution, edges derived merely based on Euclidean adjacency may include spurious connections that are not anatomically justifiable, for instance, at touching banks of sulci. By exploiting pial surface extractions, such anatomically unjustifiable connections, i.e. graph edges, are pruned out. No weight is assigned to the edges. 

\vspace{-3mm}
\subsubsection{Cortical Parcellation}
\label{sec:parcellation}
A hemisphere is parcellated into a set of regions of approximately equal volume. For satisfying equality of regional volumes, the number of regions may slightly vary between hemispheres depending on the level of volumetric asymmetry. The parcellation is performed using spectral clustering \cite{Luxburg2007}, and graph partitioner Chaco \cite{chacoUserguide} is leveraged for a computationally efficient implementation. Specifically, to parcel the left/right hemisphere in to $P$ parcels, initially, a set of vectors are defined by sampling the first $P$ Laplacian eigenvectors of the associated GHCH graph as 
\begin{align}
y_{i} 
& = [\chi_{_1}[i],\chi_{_2}[i], \ldots,\chi_{_P}[i]],  \quad i=1, \ldots, N_g^{(l)}, 
\end{align}
where $N_g^{(l)}$ denotes the number of vertices of the GHCH graph.  
Vectors $\{y_{i}\}_{i=1}^{N_{g}^{(l)}}$ are then clustered with the k-means algorithm 
in to $N$ graph vertex clusters $\{C_{j} \subset \{1, \ldots, N_g^{(l)}\} \}_{j=1}^{P}$, where $C_{1}~\cup~C_{2}~\cup~\cdots~C_{P} = \{1,\ldots, N_g^{(l)}\}$. Voxels associated to each vertex cluster $C_{i}$ are then treated as a single parcel, resulting in $N$ localized cortical parcels within the hemisphere. 

It is worth noting that in the present work the number of parcels is defined based on a specified desired resolution for the parcels across hemispheres and subjects. In other words, rather than parcellating different hemispheres all in to a fixed number of parcels, we instead keep the parcel size fix, thus allowing some variation in the number of parcels across hemispheres and subjects.        

\vspace{-3mm}
\subsubsection{Localized Cerebral Hemisphere Cortex (LCHC) Graphs}
A graph is designed for each cortical cluster, which we denote as localized cerebral hemisphere cortex (LCHC) graph. The vertex set of a LCHC graph associated to cluster $i$ consists of voxels that lie within the associated vertex cluster $C_{i}$. The edge set of the LCHC graph is defined based on the same neighbourhood connectivity principle and pruning approach as that explained in constructing GCHC graphs; in practice, the $A$ matrices of LCHC graphs can be extracted from the $A$ matrix of their associated GCHC graph. 

\vspace{-2mm}
\subsection{Spectral characterization of cortical graphs}
In the following, we define a set of spectral graph metrics that quantify morphological information across the Laplacian spectra of GHCH graphs and LCHC graphs. 

\vspace{-3mm}
\subsubsection{Spectral metric for GCHC graphs}
\label{sec:gchc-metric}
Given $N$ subjects, let $\mathcal{S}_{n}$, $n=1, \ldots, N$, denote the spectrum of the GCHC graph of the left/right hemisphere of subject $n$. At a given spectral band, denoted $\alpha - \beta$ where $\alpha \in [0,2)$, $\beta \in (0,2]$ and $\alpha < \beta$, a spectral metric is defined on the GCHC graph as
\begin{equation}
\Theta_{n}^{\alpha - \beta} = \vert s_{n}^{\alpha -\beta} \vert, 
\end{equation}
where $\vert \cdot \vert$ denotes set cardinality and set $s_{n}^{\alpha -\beta}$ is given as
\begin{align}
s_{n}^{\alpha -\beta} 
=
\begin{cases} 
\{\lambda \in \mathcal{S}_{n} \vert ~ \alpha \le \lambda \le \beta \}, \quad \alpha = 0, 
\\
\{\lambda \in \mathcal{S}_{n} \vert ~ \alpha < \lambda \le \beta \}, \quad \text{otherwise}.
\end{cases}
\end{align}

GCHC graphs, at 1 milimeter cubic resolution as presented in this work, have approximately 300 K vertices. Direct computation of $s_{n}^{\alpha -\beta}$ is thus computationally cumbersome as it requires deriving exact eigenvalues of the $L$ matrix. In particular, to compute $s_{n}^{\alpha -\beta}$ at different spectral bands spanning  the entire spectrum, a full eigendecomposition of matrix $L$ is needed, which is practically infeasible. In this work, for GCHC graphs, we compute their exact spectra within $[0, 0.1]$, i.e., lower 5\% spectral tail, and use an approximation scheme to estimate the number of eigenvalues that fall within spectral bands at upper parts of the spectra.  

The approximation is performed using the spectrum slicing method \cite[Section 3.3]{Parlett1998}, which has also been previously used in \cite{Shuman2015ieee} for approximating graph spectra. Specifically, the number of eigenvalues of $L$ that fall below a given $\alpha \in [0,2)$ can be computed as follows. Firstly, a triangular factorization of matrix $L - \alpha I$ is performed, i.e., $ L- \alpha I =  \Pi  \Delta \Pi^{T}$, where $\Pi$ is a lower triangular matrix and $\Delta$ is a diagonal matrix. Secondly, by invoking a corollary of SylvesterÕs law of inertia, it holds that the number of negative eigenvalues of $\Delta$, denoted $N_{\alpha}$, is equal to the number of negative eigenvalues of $L- \alpha I$, and thus equal to the number of eigenvalues of $L$ less than $\alpha$. Similarly, the number of eigenvalues of $L$ that fall below a given $\beta \in (0,2], \beta>\alpha$, denoted $N_{\beta}$, can be estimated. An approximation of $s_{n}^{\alpha -\beta}$ is thus given by $N_{\beta} - N_{\alpha}$.      

\vspace{-3mm}
\subsubsection{Spectral metric for LCHC graphs}
Assume the left/right hemisphere of subject $n$ being parcellated, at a desired resolution, in to $K_{n}$ parcels, thus, resulting in a set of $K_{n}$ LCHC graphs. Let $\mathcal{S}_{n,k}$, $k=1,\ldots, K_{n}$, denote the Laplacian spectrum of the $k$-th LCHC graph of subject $n$. At a given spectral band, denoted $\alpha - \beta$ where $\alpha \in [0,2)$, $\beta \in (0,2]$ and $\alpha < \beta$, a spectral metric is defined on the set of LCHC graphs as  
\begin{equation}
\theta_{n}^{\alpha-\beta} = \frac{1}{K_{n}}\sum_{k=1}^{K_{n}} \vert s_{n,k}^{\alpha -\beta} \vert, \quad n=1\ldots,N, 
\end{equation}
where set $s_{n,k}^{\alpha -\beta} $ is given as
\begin{align}
s_{n,k}^{\alpha -\beta} 
=
\begin{cases} 
\{\lambda \in \mathcal{S}_{n,k} \vert ~ \alpha \le \lambda \le \beta \}, \quad \alpha = 0, 
\\
\{\lambda \in \mathcal{S}_{n,k} \vert ~ \alpha < \lambda \le \beta \}, \quad \text{otherwise}.
\end{cases}
\end{align}
It is worth noting that $\sum_{k=1}^{K_{n}} s_{n,k}^{\alpha-\beta}$ is generally not equal to $s_{n}^{\alpha-\beta}$, nor is $\{\mathcal{S}_{n,1} \cup \mathcal{S}_{n,2} \cup \cdots \cup \mathcal{S}_{n,K_{n}}\}$ equal to $\mathcal{S}_{n}$, as has been empirically observed, consistently, across our analysis. This observation provides intuition in that it shows that the spectra of LCHC graphs can be seen as a novel decomposition of the single spectrum of the associated GCHC graph, such that the unity of the LCHC graph spectra is not equal to the single GCHC graph spectrum. Detailed theoretical analysis of this property is deferred to our future work.

\vspace{-2mm}
\section{Results}
\begin{figure}[tbp] 
   \centering
   \includegraphics[width=0.44\textwidth]{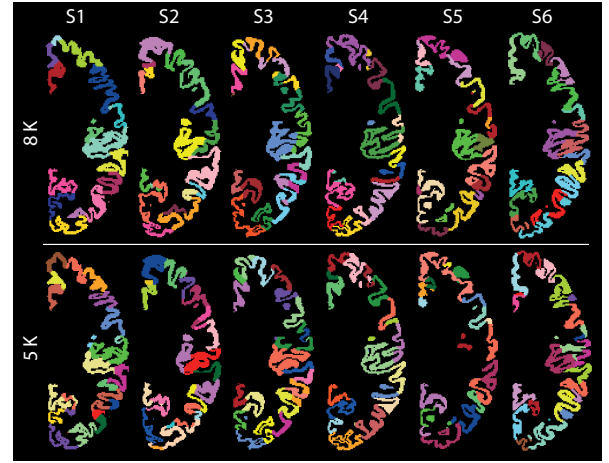}
   \caption{Parcellated right hemisphere of six subjects from the HCP database, one subject per column, at two resolutions; 40Ê$\pm$ 2 and 63 $\pm$ 3 parcels in the top and bottom rows, respectively. Slices are shown along the same MNI coordinate, at 1 mm$^{3}$ voxel resolution. Volumes of cortical regions in each hemisphere are equal, and are approximately equal across subjects; top row: $~$Ê8000 (8K) voxels, bottom row: $~$ 5000 (5K) voxels.
   The first three subjects are females and the second three are males. The use of color is to distinguish adjacent parcels and there is no link between parcels of identical color across subjects and resolutions.}
   \label{fig:regions} \vspace{-3mm}
\end{figure}

The analysis were performed on a subset of subjects from the Human Connectome Project \cite{HCP} database, consisting of 75 female and 75 male subjects, all within the age group of 31-35 years. The female and male subject subsets were selected objectively based on the numerical ordering of HCP subject identifiers, starting from smallest identifiers. Fig.~\ref{fig:regions} shows the right hemispheres of 6 of the subjects, where the first three subjects are females and the second three are males; cortical parcellations obtained using the scheme described in Section~\ref{sec:parcellation} are also illustrated. 

In the following, results from hypothesis tests on i) left-right hemisphere asymmetry and ii) hemispheric morphological differences between gender are presented. The primary objective with these tests is to study the variability of the proposed spectral metrics $\Theta_{n}^{\alpha-\beta}$ and $\theta_{n,k}^{\alpha-\beta}$, across various spectral bands and parcellation resolutions. We report p-values that result from the tests as a means to study the variations. We do not aim to attach any tag of significance to the findings, and as such, uncorrected p-values are reported.\footnote{Yet, it should be noted that a large extent of the p-values would survive even a strict Bonferroni correction for the number of spectral bands studied.} 

\begin{table}[]
\vspace{0mm} 
  \centering
\resizebox{0.49 \textwidth}{!}{
   \begin{tabular}{@{} r @{\hspace{1.3\tabcolsep}} r @{\hspace{1.3\tabcolsep}}r @{\hspace{1.3\tabcolsep}} r @{\hspace{1.3\tabcolsep}}r @{\hspace{1.3\tabcolsep}}r @{\hspace{1.3\tabcolsep}}r @{\hspace{1.3\tabcolsep}} c  @{}} 
      \toprule
      \multicolumn{1}{c}{\multirow{1}{*}{Spectral}} & \multicolumn{6}{c}{Local Graphs} & \multicolumn{1}{c}{\multirow{1}{*}{Global}}\\
      \cmidrule{2-7} 
      Band~~ & \multicolumn{1}{c}{5K~~~~} & \multicolumn{1}{c}{6K~~~~} & \multicolumn{1}{c}{7K~~~~} & \multicolumn{1}{c}{8K~~~~} & \multicolumn{1}{c}{9K~~~~} & \multicolumn{1}{c}{10K~~~~} & Graphs\\
      \midrule \relax
 0 -- 0.1
& 0.1102~~~~ & 0.1351~~~~ & 0.0500~~~~ & 0.0802~~~~ & 0.1458~~~~ & 0.1902~~~~ & 0.4179
\\
 0.1 -- 0.2
& 0.0971~~~~ & 0.1396~~~~ & 0.0527~~~~ & 0.0577~~~~ & 0.1796~~~~ & 0.1960~~~~ &   0.0320
\\
 0.2 -- 0.3
& 0.2107~~~~ & 0.4136~~~~ & 0.0985~~~~ & 0.1896~~~~ & 0.6185~~~~ & 0.9880~~~~ &   0.0012
\\
 0.3 -- 0.4
& 0.3435~~~~ & 0.9640~~~~ & 0.3571~~~~ & 0.5058~~~~ & 0.7881~~~~ & 0.4511~~~~ &   {4.8  {$\times 10^{-4~}$}}
 \\
 0.4 -- 0.5
& 0.8983~~~~ & 0.5413~~~~ & 0.5627~~~~ & 0.7224~~~~ & 0.5798~~~~ & 0.0594~~~~ &   {4.9  {$\times 10^{-4~}$}}
 \\
 0.5 -- 0.6
& 0.8318~~~~ & 0.6903~~~~ & 0.3551~~~~ & 0.7852~~~~ & 0.4009~~~~ & 0.0650~~~~ &   {3.7  {$\times 10^{-4~}$}}
 \\
 0.6 -- 0.7
& 0.1527~~~~ & 0.9281~~~~ & 0.1381~~~~ & 0.4158~~~~ & 0.9236~~~~ & 0.2635~~~~ &   {6.8  {$\times 10^{-4~}$}}
 \\
 0.7 -- 0.8
& 0.0432~~~~ & 0.5602~~~~ & 0.0694~~~~ & 0.1689~~~~ & 0.8318~~~~ & 0.4288~~~~ &   {7.1  {$\times 10^{-4~}$}}
 \\
 0.8 -- 0.9
& 0.0350~~~~ & 0.3801~~~~ & 0.0610~~~~ & 0.2627~~~~ & 0.6264~~~~ & 0.3863~~~~ &   0.0014
 \\
 0.9 -- 1
& 0.9356~~~~ & 0.3650~~~~ & 0.9461~~~~ & 0.7042~~~~ & 0.4009~~~~ & 0.0666~~~~ &   0.0015
 \\
 1 -- 1.1
& 0.3229~~~~ & 0.0754~~~~ & 0.3822~~~~ & 0.3192~~~~ & 0.1044~~~~ & 0.0153~~~~ &   {~7  {$\times 10^{-4~}$}}
 \\
 1.1 -- 1.2
& 0.0155~~~~ & 0.1036~~~~ & 0.0112~~~~ & 0.0360~~~~ & 0.2013~~~~ & 0.5767~~~~ &   0.0025
 \\
 1.2 -- 1.3
& 0.0237~~~~ & 0.1852~~~~ & 0.0104~~~~ & 0.0430~~~~ & 0.3397~~~~ & 0.9341~~~~ &   0.0017
 \\
 1.3 -- 1.4
& 0.1411~~~~ & 0.1463~~~~ & 0.0584~~~~ & 0.0821~~~~ & 0.2524~~~~ & 0.2926~~~~ &   0.0220
 \\
 1.4 -- 1.5
&   {0.0197}~~~~ &   {0.0221}~~~~ &   {0.0409}~~~~ &   {0.0103}~~~~ &   {0.0300}~~~~ &   {0.0147}~~~~ &   0.0010 
 \\
 1.5 -- 1.6
&   {2.3  {$\times 10^{-7~}$}} &   8.2  {$\times 10^{-6~}$} &   {1.2  {$\times 10^{-8~}$}} &   {~5  {$\times 10^{-8~}$}} &  {1.3 {$\times 10^{-7~}$}} &   {7.3  {$\times 10^{-9~}$}} &   {~3.8  {$\times 10^{-9~}$}}
 \\
 1.6 -- 1.7
&   {0.0112}~~~~ &   {9.6  {$\times 10^{-4~}$}} &   {1.6 {$\times 10^{-4~}$}} &   {0.0012}~~~ &   {0.0025}~~~~ &   {1.4 {$\times 10^{-4~}$}} &   {~1  {$\times 10^{-5~}$}}
 \\
 1.7 -- 1.8
& {0.0017}~~~~ & {0.0124}~~~~ &   {0.0013}~~~~ & {6.8  {$\times 10^{-5~}$}} &   {0.0180}~~~~ &   {4 {$\times 10^{-4~}$}} &   {1.2  {$\times 10^{-4~}$}}
 \\
 1.8 -- 1.9
& 0.5537~~~~ &   0.3065~~~~ & 0.3386~~~~ & 0.3348~~~~ & 0.0324~~~~ & 0.1797~~~~ &   0.0694
 \\
 1.9 -- 2
& 0.2311~~~~ & 0.9179~~~~ & 0.1056~~~~ & 0.7198~~~~ & 0.7172~~~~ & 0.0684~~~~ & 0.4057
 \\
      \bottomrule
   \end{tabular}
} 
\caption{Hemispheric asymmetry. P-values from Wilcoxon rank-sum tests on sets i) $\{\theta_{n}^{\alpha - \beta}\}_{n=1,\ldots,75}$ for the left hemispheres of the male group, and ii) $\{\theta_{n}^{\alpha - \beta}\}_{n=1,\ldots,75}$ for the right hemispheres of the male group, across different spectral bands and parcellation resolutions. Similarly, the last columns shows p-values from Wilcoxon rank-sum tests on sets i) $\{\Theta_{n}^{\alpha - \beta}\}_{n=1,\ldots,75}$ for the left hemispheres of the male group, and ii) $\{\Theta_{n}^{\alpha - \beta}\}_{n=1,\ldots,75}$ for the right hemispheres of the male group. The same tests performed on the female group led to similar results; results not presented due to limit of space.} 
\label{tab:pvals-LR}
\vspace{-3mm}
\end{table}

\vspace{-3mm}
\subsection{Graph spectral markers of hemispheric asymmetry}
\label{sec:asymmetry}
\begin{figure}[t]    \centering
   \includegraphics[width=0.42\textwidth]{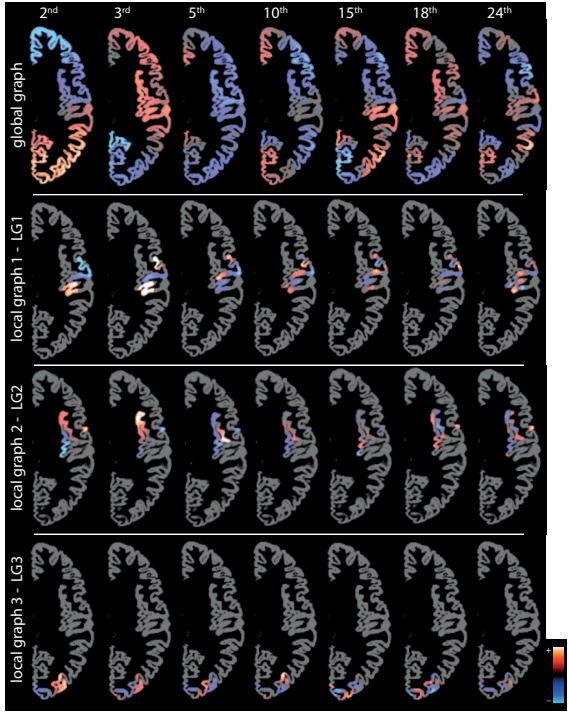} 
   \caption{Eigenvectors of the GCHC graph and three LCHC graphs associated to the right hemisphere of subject S1, cf. Fig.~\ref{fig:regions}, overlaid on the cortical ribbon shown in gray. Note that the eigenvectors are defined in 3D space whereas only a single axial slice of them is shown, which limits manifesting their full spatial variation. Eigenvectors of LCHC graphs better capture localized morphological variations, whereas those of the GCHC graph better capture global topological variations.
}
   \label{fig:eigenvectors}
   \vspace{-3mm}
\end{figure}

Variations in the cortical morphology between left and right hemisphere has been numerously reported in literature, see for example \cite{Kong2018}. Wilcoxon rank-sum test analysis was performed on the group of left and right hemispheres, implemented separately for each gender to prevent bias. The tests were performed on $\Theta_{n}^{\alpha-\beta}$ and $\theta_{n,k}^{\alpha-\beta}$ metrics across different spectral bands. Table~\ref{tab:pvals-LR} summarizes the resulting statistical p-values. P-values obtained from tests on $\Theta_{n}^{\alpha-\beta}$Êare lower than corresponding ones obtained from test on $\theta_{n}^{\alpha-\beta}$, across different parcellation resolutions, excluding the first spectral band. This suggests the superiority of GCHC graphs in encoding hemispheric asymmetry over LCHC graphs. Interestingly, this observation may be related to the strong fronto-occipital asymmetry pattern in cortical thickness as reported in \cite{Kong2018}, which itself has been suggested to be related to the Yakovlevian torque, an overall hemispheric twist giving rise to the frontal and occipital petalia. That is, eigenvectors that encode the global structure of the hemisphere, i.e., eigenvectors of GCHC graphs, can better capture this elongated pattern of asymmetry compared to eigenvectors which have localized support, i.e., those of LCHC graphs. Fig.~\ref{fig:eigenvectors} shows eigenvectors of the Laplacian matrices of three LCHC graphs and the associated GCHC graph. Eigenvectors associated to smaller eigenvalues represent slower spatial harmonics, whereas those associated to higher eigenvalues, encode more subtle spatial patterns.

\begin{figure}[] 
   \centering
   \includegraphics[width=0.49\textwidth]{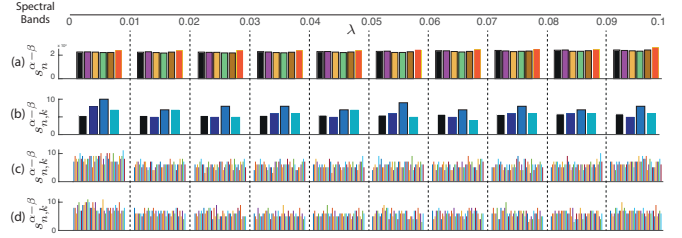} 
 \caption{(a)-(d) Distribution of graph Laplacian eigenvalues in the lower-end spectra of cerebral hemisphere cortex graphs: (a) GCHC graphs associated to the six right hemispheres shown in Fig.~\ref{fig:regions}. (b) GCHC graph and the three LCHC graphs associated to the hemisphere shown in Fig.~\ref{fig:eigenvectors}. The black bars in (b) are obtained by dividing the black bars in (a) by the number of local graphs in the hemisphere, $N = 42$. (c) 42 LCHC graphs associated to the hemisphere shown in Fig.~\ref{fig:eigenvectors}. (d) The same as in (c) but for the subject's left hemisphere, $N = 41$. 
In (b)-(d), local graphs have size 8K.} 
   \label{fig:merged}
\end{figure}

The results further show that both metrics $\Theta_{n,k}^{\alpha-\beta}$ and $\theta_{n,k}^{\alpha-\beta}$Êexhibit particularly significant variation between left and right hemispheres within spectral range $[1.4, 1.8]$. Moreover, at spectral band $[0, 0.1]$, p-values associated to LCHC graphs show lower values than that associated to the GCHC graph. This observation, together with noting that eigenvectors associated to the lower end of the spectrum provide more information on the macro-scale structure of cortical hemispheres, suggests further scrutiny of this spectral band. Fig.~\ref{fig:merged} provides visual intuition on distribution of eigenvalues at 10 sub-bands within spectral band $[0,0.1]$, showing the extent of spectral variation across LCHC graphs compared to the associated GCHC graph. {Wilcoxon rank-sum tests analysis were then performed on these 10 sub-bands, see Table~\ref{tab:pvals-LR-lowerend}, showing that a greater level of hemispheric asymmetry can be manifested at a number of these narrow spectral bands; see underlined values in Table~\ref{tab:pvals-LR-lowerend}.}

\begin{table}[]
\vspace{0mm} 
  \centering
\resizebox{0.49 \textwidth}{!}{
   \begin{tabular}{@{} r @{\hspace{1.3\tabcolsep}} r @{\hspace{1.3\tabcolsep}}r @{\hspace{1.3\tabcolsep}} r @{\hspace{1.3\tabcolsep}}r @{\hspace{1.3\tabcolsep}}r @{\hspace{1.3\tabcolsep}}r @{\hspace{1.3\tabcolsep}} c  @{}} 
      \toprule
      \multicolumn{1}{c}{\multirow{1}{*}{Spectral}} & \multicolumn{6}{c}{Local Graphs} & \multicolumn{1}{c}{\multirow{1}{*}{Global}}\\
      \cmidrule{2-7} 
      Band~~ & \multicolumn{1}{c}{5K~~~~~~~~} & \multicolumn{1}{c}{6K~~~~~~~~} & \multicolumn{1}{c}{7K~~~~~~~~} & \multicolumn{1}{c}{8K~~~~~~~~} & \multicolumn{1}{c}{9K~~~~~~~~} & \multicolumn{1}{c}{10K~~~~~~~~} & Graphs\\
      \midrule \relax
       0 -- 0.1~~
& 0.1102~~~~ & 0.1351~~~~ & 0.0500~~~~ & 0.0802~~~~ & 0.1458~~~~ & 0.1902~~~~ & 0.4179
\\
\cmidrule{1-8} \relax
0 -- 0.01
& 0.3229~~~~ & 0.4670~~~~ & 0.3210~~~~ & 0.5538~~~~ & 0.2290~~~~ & 0.9730~~~~ & 0.4800
\\ 
0.01 -- 0.02
& 0.1153~~~~ & 0.4821~~~~ &0.1004~~~~ & \underline{0.0317}~~~~ & 0.3415~~~~ & \underline{0.0633}~~~~ &0.4590      
\\  
0.02 -- 0.03
&  0.4705~~~~ & \underline{0.0149}~~~~ &\underline{0.0429}~~~~ &0.2297~~~~ &  0.2092~~~~ &0.3302~~~~ &0.7420
\\  
0.03 -- 0.04
&  \underline{0.0059}~~~~ & 0.5388~~~~ &\underline{0.0202}~~~~ &0.1499~~~~ &0.2275~~~~ &0.4774~~~~ &\underline{0.3270}
\\  
0.04 -- 0.05
& 0.5526~~~~ & 0.3425~~~~ & 0.2508~~~~ & 0.2225~~~~ & \underline{0.0586}~~~~ & \underline{0.0679}~~~~	& 0.6280
\\ 
0.05 -- 0.06
& 0.2823~~~~ & \underline{0.0291}~~~~ & 0.2386~~~~ &\underline{0.0173}~~~~ & 0.2462~~~~ & 0.9745~~~~ & 0.9520
\\  
0.06 -- 0.07
& \underline{0.0312}~~~~ & 0.3065~~~~ &\underline{0.0337}~~~~ &0.2363~~~~ & 0.2106~~~~ &\underline{0.0237}~~~~ &\underline{0.0770}
\\  
0.07 -- 0.08
& 0.1270~~~~ & \underline{0.1080}~~~~ & 0.4211~~~~ & 0.2492~~~~ & \underline{0.1421}~~~~ & 0.8760~~~~ & \underline{0.3756}
\\  
0.08 -- 0.09
& 0.6384~~~~ & 0.8539~~~~ & 0.1973~~~~ &  \underline{0.0424}~~~~ & 0.4844~~~~ &  \underline{0.0738}~~~~ & 0.4780
\\  
0.09 -- 0.1~~
& \underline{0.0947}~~~~ & 0.3110~~~~ & \underline{0.0291}~~~~ & 0.1473~~~~ & \underline{0.0437}~~~~ & 0.5626~~~~ & \underline{0.3596}
 \\
      \bottomrule
   \end{tabular}
} 
\caption{Same as in Table~\ref{tab:pvals-LR} but on a set of narrower spectral bands spanning spectral range $[0, 0.1]$; results on spectral band 0 -- 0.1 replicated from Table~\ref{tab:pvals-LR}. Tests that led to p-values lower than that obtained on corresponding tests at spectral band $[0,0.1]$ are underlined.} 
\label{tab:pvals-LR-lowerend}
\vspace{-3mm}
\end{table}

\vspace{-3mm}
\subsection{Cortical graph spectral markers of gender}
Variations in the cortical morphology of male and female subjects has been suggested in many studies, see for example \cite{Ritchie2018}. Wilcoxon rank-sum test analysis was performed on the groups of male and females subjects. The test was implemented separately for each hemisphere to prevent bias due to hemispheric asymmetry. The tests were performed on metrics $\Theta_{n}^{\alpha-\beta}$ and $\theta_{n,k}^{\alpha-\beta}$ across different spectral bands. The resulting p-values are shown in Table~\ref{tab:pvals-sex}. All tests on $\theta_{n,k}^{\alpha-\beta}$ led to lower p-values relative to the associated test on $\Theta_{n}^{\alpha-\beta}$, suggesting the superiority of LCHC graphs over GCHC graphs in collectively providing a more discriminative encoding of cortical morphology across gender. Interestingly, this result is in contrast to that observed on tests of hemispheric asymmetry where the global metric $\Theta_{n}^{\alpha-\beta}$ was found to provide better discrimination. This observation can be interpreted as that discrimination of gender is best exhibited as regional variations in cortical morphology rather than as global hemispheric variations.
 
With similar reasoning as that provided in Section~\ref{sec:asymmetry}, the tests were also performed at narrow spectral bands in the lower-end of the spectra. The results are shown in Table~\ref{tab:pvals-sex-lowerend}. Table~\ref{tab:pvals-sex}, all tests led to significant p-values. 
In contrast to tests on hemispheric asymmetry, these tests on gender variation at narrow spectral bands in the lower end of the spectra resulted in few lower p-values than that obtained on the spectral band $[0,0.1]$; see underlined values in Table~\ref{tab:pvals-sex-lowerend}. 

\begin{table*}[h]
   \centering
   \vspace{0mm} 
\resizebox{0.94 \textwidth}{!}{
   \begin{tabular}{@{} r  @{\hspace{1.3\tabcolsep}} r @{\hspace{1.3\tabcolsep}}r @{\hspace{1.3\tabcolsep}} r @{\hspace{1.3\tabcolsep}}r @{\hspace{1.3\tabcolsep}}r @{\hspace{1.3\tabcolsep}}r @{\hspace{1.3\tabcolsep}} c  @{} @{\hspace{1.3\tabcolsep}} r @{\hspace{1.3\tabcolsep}}r @{\hspace{1.3\tabcolsep}} r @{\hspace{1.3\tabcolsep}}r @{\hspace{1.3\tabcolsep}}r @{\hspace{1.3\tabcolsep}}r @{\hspace{1.3\tabcolsep}} c  @{}} 
      \toprule
      \multicolumn{1}{c}{\multirow{1}{*}{~}} & \multicolumn{7}{c}{Left Hemisphere} & \multicolumn{7}{c}{Right Hemisphere} 
      \\
      \cmidrule(lr){2-8} \cmidrule(lr){9-15} 
      \multicolumn{1}{c}{\multirow{1}{*}{Spectral}} & \multicolumn{6}{c}{Local Graphs} & \multicolumn{1}{c}{\multirow{1}{*}{Global}} & \multicolumn{6}{c}{Local Graphs} & \multicolumn{1}{c}{\multirow{1}{*}{Global}}
      \\
      \cmidrule(lr){2-7} 
      \cmidrule(lr){9-14}
      Band~~ & \multicolumn{1}{c}{5K~~~~} & \multicolumn{1}{c}{6K~~~~} & \multicolumn{1}{c}{7K~~~~} & \multicolumn{1}{c}{8K~~~~} & \multicolumn{1}{c}{9K~~~~} & \multicolumn{1}{c}{10K~~~~} & Graphs & \multicolumn{1}{c}{5K~~~~} & \multicolumn{1}{c}{6K~~~~} & \multicolumn{1}{c}{7K~~~~} & \multicolumn{1}{c}{8K~~~~} & \multicolumn{1}{c}{9K~~~~} & \multicolumn{1}{c}{10K~~~~} & Graphs
      \\
\cmidrule(lr){1-1} \cmidrule(lr){2-8} \cmidrule(lr){9-15} 
~0 -- 0.1 
& 7.3 {$\times 10^{-14}$}
& 4.8 {$\times 10^{-14}$}  
& 5.3 {$\times 10^{-14}$} 
& 7.1 {$\times 10^{-13}$} 
& ~1 {$\times 10^{-13}$}
& 1.1 {$\times 10^{-12}$} 
& ~2 {$\times 10^{-9~}$} 
& 1.4 {$\times 10^{-12}$}
& 8.5 {$\times 10^{-13}$}  
& 5.3 {$\times 10^{-13}$} 
& 7.9 {$\times 10^{-13}$} 
& 6.1 {$\times 10^{-12}$}
& 6.3 {$\times 10^{-13}$} 
& 7.2 {$\times 10^{-10}$} 
\\
0.1 -- 0.2 
& 9.5 {$\times 10^{-14}$}
& 6.6 {$\times 10^{-14}$}  
& ~4 {$\times 10^{-13}$} 
& ~2 {$\times 10^{-12}$} 
& 2.4 {$\times 10^{-13}$}
& 3.7 {$\times 10^{-12}$} 
& {4.4 $\times$ $10$}$^{-4~}$ 
& 8.3 {$\times 10^{-12}$}
& ~2 {$\times 10^{-12}$}  
& 1.2 {$\times 10^{-12}$} 
& 3.9 {$\times 10^{-12}$} 
& 3.7 {$\times 10^{-11}$}
& 4.7 {$\times 10^{-12}$} 
& 7.3 {$\times 10^{-5~}$}
\\
0.2 -- 0.3 
& 7.8 {$\times 10^{-12}$}
& 9.7 {$\times 10^{-12}$}  
& 1.5 {$\times 10^{-11}$} 
& 4.6 {$\times 10^{-9~}$} 
& 3.8 {$\times 10^{-10}$}
& 1.1 {$\times 10^{-7~}$} 
& {0.9371}~~~~ 
& 7.1 {$\times 10^{-9~}$}
& ~3 {$\times 10^{-10}$}  
& 2.9 {$\times 10^{-10}$} 
& 2.2 {$\times 10^{-8~}$} 
& 2.2 {$\times 10^{-8~}$}
& ~3 {$\times 10^{-9~}$} 
& 0.8553~~~~ 
\\
0.3 -- 0.4 
& ~7{$\times 10^{-4~}$}
& 6.8 {$\times 10^{-4~}$}  
& 4.2 {$\times 10^{-5~}$} 
& 0.0208~~~~ 
& 0.0137~~~~ 
& 0.0101~~~~ 
& {0.0098}~~~~ 
& 2.1 {$\times 10^{-4~}$}
& 1.2 {$\times 10^{-5~}$}  
& 1.1 {$\times 10^{-4~}$} 
& 0.0587~~~~ 
& 4.8 {$\times 10^{-4~}$}
& 0.0188~~~~ 
& 0.0311~~~~
\\
0.4 -- 0.5 
& 0.1210~~~~ 
& 0.0361~~~~  
& 0.0044~~~~ 
& 0.8216~~~~  
& 0.3165~~~~ 
& 0.4613~~~~ 
& {6.4 $\times$ $10$}$^{-4~}$ 
& 0.4556~~~~ 
& 0.0308~~~~ 
& 0.0869~~~~
& 0.5896~~~~
& 0.0264~~~~
& 0.1796~~~~
& 0.0055~~~~
\\
0.5 -- 0.6 
& 0.0337~~~~ 
& 0.0214~~~~  
& 0.0031~~~~  
& 0.5376~~~~  
& 0.3650~~~~ 
& 0.3956~~~~  
& {0.0012}~~~~  
& 0.0070~~~~
& 0.0081~~~~
& 0.0074~~~~
& 0.6506~~~~
& 0.0095~~~~
& 0.0978~~~~
& 0.0063~~~~ 
\\
0.6 -- 0.7 
& ~4 {$\times 10^{-6~}$}
& 3.3 {$\times 10^{-4~}$}  
& 4.5 {$\times 10^{-6~}$} 
& 0.0316~~~~  
& 0.0126~~~~ 
& 0.0287~~~~  
& {0.0036}~~~~ 
& ~7 {$\times 10^{-4~}$}
& 7.9 {$\times 10^{-5~}$}  
& 0.0017~~~~ 
& 0.1442~~~~ 
& 0.0031~~~~ 
& 0.0164~~~~ 
& 0.0121~~~~ 
\\
0.7 -- 0.8 
& 1.1 {$\times 10^{-5~}$}
& ~2 {$\times 10^{-4~}$}  
& 1.4 {$\times 10^{-5~}$} 
& 0.0430~~~~ 
& 0.0232~~~~ 
& 0.0562~~~~  
& {0.0022}~~~~ 
& 0.0013~~~~ 
& 9.3 {$\times 10^{-6~}$} 
& 8.6 {$\times 10^{-4~}$} 
& 0.2073~~~~
& 0.0020~~~~
& 0.0300~~~~ 
& 0.0092~~~~ 
\\
0.8 -- 0.9 
& ~8 {$\times 10^{-4~}$}
& 0.0023~~~~ 
& 1.9 {$\times 10^{-4~}$} 
& 0.2799~~~~  
& 0.0805~~~~ 
& 0.1473~~~~  
& {0.0016}~~~~ 
& 0.0150~~~~
& 7.3 {$\times 10^{-4~}$}  
& 0.0027~~~~
& 0.3591~~~~
& 0.0077~~~~
& 0.0511~~~~
& 0.0073~~~~ 
\\
0.9 -- 1~~~ 
& 1.1 {$\times 10^{-8~}$}
& 5.5 {$\times 10^{-8~}$}  
& 1.8 {$\times 10^{-6~}$} 
& 4.4 {$\times 10^{-8~}$} 
& 1.3 {$\times 10^{-5~}$}
& 3.8 {$\times 10^{-7}$} 
& 6.8 {$\times 10^{-6~}$} 
& 2.3 {$\times 10^{-9~}$}
& 1.3 {$\times 10^{-6~}$}  
& 4.7 {$\times 10^{-6~}$} 
& ~1 {$\times 10^{-6~}$} 
& 1.1 {$\times 10^{-5~}$}
& 1.7 {$\times 10^{-5~}$} 
& 7.7 {$\times 10^{-5~}$}
\\
~1 -- 1.1 
& 2.9 {$\times 10^{-10}$}
& 1.5 {$\times 10^{-9~}$}  
& 1.3 {$\times 10^{-8~}$} 
& 2.1 {$\times 10^{-9~}$} 
& 5.5 {$\times 10^{-7~}$}
& 7.6 {$\times 10^{-9~}$} 
& 1.7 {$\times 10^{-6~}$} 
& 2.1 {$\times 10^{-10}$}
& 1.3 {$\times 10^{-7~}$}  
& 4.3 {$\times 10^{-7~}$} 
& 8.9 {$\times 10^{-8~}$} 
& 6.6 {$\times 10^{-7~}$}
& 1.9 {$\times 10^{-6~}$} 
& 3.1 {$\times 10^{-5~}$}
\\
1.1 -- 1.2 
& 5.5 {$\times 10^{-13}$}
& 3.9 {$\times 10^{-12}$}  
& 6.7 {$\times 10^{-12}$} 
& 4.6 {$\times 10^{-9~}$} 
& 1.3 {$\times 10^{-9~}$}
& 9.6 {$\times 10^{-8~}$} 
& 0.2824~~~~  
& 7.8 {$\times 10^{-10}$}
& 1.1 {$\times 10^{-11}$}  
& ~1 {$\times 10^{-10}$} 
& 9.8 {$\times 10^{-9~}$} 
& 2.6 {$\times 10^{-8~}$}
& 1.3 {$\times 10^{-8~}$} 
& 0.4354~~~~
\\
1.2 -- 1.3
& 8.3 {$\times 10^{-11}$}
& 8.7 {$\times 10^{-10}$}  
& 3.3 {$\times 10^{-10}$} 
& 8.5 {$\times 10^{-7~}$} 
& 1.4 {$\times 10^{-6~}$}
& 1.4 {$\times 10^{-5~}$} 
& 0.0641~~~~ 
& 7.3 {$\times 10^{-9~}$}
& 2.2 {$\times 10^{-10}$}  
& 5.1 {$\times 10^{-9~}$} 
& 6.1 {$\times 10^{-6~}$} 
& 2.3 {$\times 10^{-7~}$}
& 1.2 {$\times 10^{-6~}$} 
& 0.1463~~~~ 
\\
1.3 -- 1.4 
& 7.1 {$\times 10^{-14}$}
& 3.1 {$\times 10^{-14}$}  
& 1.5 {$\times 10^{-13}$} 
& 1.3 {$\times 10^{-12}$} 
& 2.1 {$\times 10^{-13}$}
& 5.2 {$\times 10^{-12}$} 
& ~9 {$\times 10^{-4~}$} 
& 4.3 {$\times 10^{-12}$}
& 5.8 {$\times 10^{-13}$}  
& 5.2 {$\times 10^{-13}$} 
& 2.1 {$\times 10^{-12}$} 
& 2.2 {$\times 10^{-11}$}
& 1.2 {$\times 10^{-12}$} 
& 8.8 {$\times 10^{-5~}$} 
\\
1.4 -- 1.5
& 1.4 {$\times 10^{-8~}$}
& 4.2 {$\times 10^{-9~}$}  
& 4.1 {$\times 10^{-9~}$} 
& 2.4 {$\times 10^{-9~}$} 
& 4.4 {$\times 10^{-9~}$}
& 7.8 {$\times 10^{-9~}$} 
& 7.7 {$\times 10^{-9~}$} 
& 6.9 {$\times 10^{-8~}$}
& 1.8 {$\times 10^{-6~}$}  
& 1.2 {$\times 10^{-7~}$} 
& 3.6 {$\times 10^{-7~}$} 
& 7.1 {$\times 10^{-7~}$}
& ~3 {$\times 10^{-7~}$} 
& 6.3 {$\times 10^{-7~}$} 
\\
1.5 -- 1.6 
& 0.0080~~~~ 
& 0.0354~~~~   
& 0.0071~~~~ 
& 0.0021~~~~  
& 0.0149~~~~ 
& 0.0279~~~~ 
& 0.0517~~~~ 
& 0.0100~~~~
& 0.0461~~~~
& 0.0026~~~~
& 0.0372~~~~
& 0.0561~~~~
& 0.1446~~~~
& 0.0482~~~~
\\
1.6 -- 1.7 
& 0.0274~~~~ 
& 0.0208~~~~ 
& 0.0286~~~~ 
& 0.0073~~~~ 
& 0.1634~~~~ 
& 0.1553~~~~ 
& 0.0070~~~~ 
& 0.1370~~~~
& 0.1880~~~~
& 0.0596~~~~
& 0.0381~~~~ 
& 0.0829~~~~
& 0.0525~~~~
& 0.0234~~~~
\\
1.7 -- 1.8 
& 0.6470~~~~ 
& 0.5400~~~~ 
& 0.3630~~~~ 
& 0.5600~~~~ 
& 0.3569~~~~ 
& 0.6562~~~~ 
& 0.7728~~~~ 
& 0.0340~~~~
& 0.1089~~~~
& 0.0816~~~~
& 0.0817~~~~
& 0.5801~~~~
& 0.1973~~~~
& 0.3025~~~~
\\
1.8 -- 1.9 
& 0.0953~~~~ 
& 0.0274~~~~ 
& 0.8330~~~~ 
& 0.4340~~~~ 
& 0.2237~~~~ 
& 0.3012~~~~ 
& 0.1161~~~~ 
& 0.7029~~~~
& 0.4842~~~~
& 0.7187~~~~
& 0.3276~~~~
& 0.0756~~~~
& 0.4850~~~~
& 0.5930~~~~
\\
1.9 -- 2~~~ 
& 0.2130~~~~ 
& 0.6999~~~~ 
& 0.6520~~~~ 
& 0.6490~~~~ 
& 0.5055~~~~ 
& 0.2219~~~~ 
& 0.1936~~~~ 
& 0.0883~~~~
& 0.8214~~~~
& 0.0323~~~~
& 0.3098~~~~
& 0.2761~~~~
& 0.7508~~~~
& 0.5836~~~~
\\
\\
      \bottomrule
   \end{tabular}
} 
\caption{Validation of LCHC and GCHC graph spectral metrics for discrimination of gender. P-values from Wilcoxon rank-sum tests on $\{\theta_{n}^{\alpha - \beta}\}_{n=1,\ldots,75}$ on groups: i) the set of 75 left/right hemispheres of the male group, and ii) the set of 75 left/right hemispheres of the female group, are presented. Similarly, p-values from Wilcoxon rank-sum tests on $\{\Theta_{n}^{\alpha - \beta}\}_{n=1,\ldots,75}$ on the same two groups are also presented. For both LCHC and GCHC graphs, tests were performed across different spectral bands, and for the LCHC graphs, also across different parcellation resolutions.} 
\label{tab:pvals-sex}
\vspace{-3mm}
\end{table*}

\begin{table*}[h]
   \centering
   \vspace{0mm} 
\resizebox{0.94 \textwidth}{!}{
   \begin{tabular}{@{} r  @{\hspace{1.3\tabcolsep}} r @{\hspace{1.3\tabcolsep}}r @{\hspace{1.3\tabcolsep}} r @{\hspace{1.3\tabcolsep}}r @{\hspace{1.3\tabcolsep}}r @{\hspace{1.3\tabcolsep}}r @{\hspace{1.3\tabcolsep}} c  @{} @{\hspace{1.3\tabcolsep}} r @{\hspace{1.3\tabcolsep}}r @{\hspace{1.3\tabcolsep}} r @{\hspace{1.3\tabcolsep}}r @{\hspace{1.3\tabcolsep}}r @{\hspace{1.3\tabcolsep}}r @{\hspace{1.3\tabcolsep}} c  @{}} 
      \toprule
      \multicolumn{1}{c}{\multirow{1}{*}{~}} & \multicolumn{7}{c}{Left Hemisphere} & \multicolumn{7}{c}{Right Hemisphere} 
      \\
      \cmidrule(lr){2-8} \cmidrule(lr){9-15} 
      \multicolumn{1}{c}{\multirow{1}{*}{Spectral}} & \multicolumn{6}{c}{Local Graphs} & \multicolumn{1}{c}{\multirow{1}{*}{Global}} & \multicolumn{6}{c}{Local Graphs} & \multicolumn{1}{c}{\multirow{1}{*}{Global}}
      \\
      \cmidrule(lr){2-7} 
      \cmidrule(lr){9-14}
      Band~~ & \multicolumn{1}{c}{5K~~~~} & \multicolumn{1}{c}{6K~~~~} & \multicolumn{1}{c}{7K~~~~} & \multicolumn{1}{c}{8K~~~~} & \multicolumn{1}{c}{9K~~~~} & \multicolumn{1}{c}{10K~~~~} & Graphs & \multicolumn{1}{c}{5K~~~~} & \multicolumn{1}{c}{6K~~~~} & \multicolumn{1}{c}{7K~~~~} & \multicolumn{1}{c}{8K~~~~} & \multicolumn{1}{c}{9K~~~~} & \multicolumn{1}{c}{10K~~~~} & Graphs
      \\
\cmidrule(lr){1-1} \cmidrule(lr){2-8} \cmidrule(lr){9-15} 
~0 -- 0.1 
& 7.3 {$\times 10^{-14}$}
& 4.8 {$\times 10^{-14}$}  
& 5.3 {$\times 10^{-14}$} 
& 7.1 {$\times 10^{-13}$} 
& ~1 {$\times 10^{-13}$}
& 1.1 {$\times 10^{-12}$} 
& ~2 {$\times 10^{-9~}$} 
& 1.4 {$\times 10^{-12}$}
& 8.5 {$\times 10^{-13}$}  
& 5.3 {$\times 10^{-13}$} 
& 7.9 {$\times 10^{-13}$} 
& 6.1 {$\times 10^{-12}$}
& 6.3 {$\times 10^{-13}$} 
& 7.2 {$\times 10^{-10}$} 
\\
\cmidrule{1-15} \relax
0 -- 0.01
      & 3.5 {$\times 10^{-12}$}
      & 4.3 {$\times 10^{-12}$}  
      & 1.8 {$\times 10^{-11}$} 
      & 3.4 {$\times 10^{-10}$} 
      & 2.1 {$\times 10^{-12}$}
      & 4.5 {$\times 10^{-11}$} 
      & \underline{3.1 {$\times 10^{-10}$}} 
& 2.4 {$\times 10^{-11}$}
      & 2.6 {$\times 10^{-11}$}  
      & 1.3 {$\times 10^{-11}$} 
      & 1.9 {$\times 10^{-12}$} 
      & 4.2 {$\times 10^{-10}$}
      & 7.2 {$\times 10^{-13}$} 
      & 6.1 {$\times 10^{-9~}$}  
\\
      0.01 -- 0.02
      & 6.4 {$\times 10^{-12}$}        
      & 3.5 {$\times 10^{-10}$} 
      & 3.1 {$\times 10^{-12}$} 
      & \underline{3.2 {$\times 10^{-13}$}} 
      & 1.7 {$\times 10^{-12}$}
      & 1.8 {$\times 10^{-10}$} 
      & 1.7 {$\times 10^{-8~}$} 
      & 2.1 {$\times 10^{-10}$}        
      & 1.1 {$\times 10^{-10}$} 
      & 1.8 {$\times 10^{-12}$} 
      & 1.3 {$\times 10^{-9~}$} 
      & \underline{~6 {$\times 10^{-13}$}}
      & 6.6 {$\times 10^{-10}$} 
      & \underline{2.6 {$\times 10^{-10}$}}
 \\
   0.02 -- 0.03 
      & 1.3 {$\times 10^{-9~}$} 
      & 1.8 {$\times 10^{-13}$} 
      & ~9 {$\times 10^{-11}$}  
      & 1.2 {$\times 10^{-10}$} 
      & 2.4 {$\times 10^{-11}$} 
      & ~1 {$\times 10^{-11}$} 
      & \underline{1.1 {$\times 10^{-9~}$}}  
& 4.5 {$\times 10^{-11}$} 
      & 2.7 {$\times 10^{-10}$} 
      & 6.6 {$\times 10^{-10}$}  
      & 5.3 {$\times 10^{-12}$} 
      & 3.6 {$\times 10^{-11}$} 
      & ~2 {$\times 10^{-12}$} 
      & 5.6 {$\times 10^{-9~}$}
\\
      0.03 -- 0.04
      & 7.6 {$\times 10^{-12}$} 
      & 7.1 {$\times 10^{-11}$} 
      & 3.1 {$\times 10^{-13}$} 
      & 3.7 {$\times 10^{-10}$}
      & 4.6 {$\times 10^{-11}$}
      & 4.1 {$\times 10^{-10}$}
      & 2.8 {$\times 10^{-9~}$}
      & 5.3 {$\times 10^{-11}$} 
      & 1.3 {$\times 10^{-10}$} 
      & 4.5 {$\times 10^{-10}$} 
      & 4.9 {$\times 10^{-10}$}
      & 7.5 {$\times 10^{-11}$}
      & 5.8 {$\times 10^{-11}$}
      & \underline{4.4 {$\times 10^{-10}$}}
\\
0.04 -- 0.05 
& 2.7 {$\times 10^{-10}$}
& 7.2 {$\times 10^{-9~}$}
& 2.1 {$\times 10^{-10}$}
& 4.2 {$\times 10^{-10}$}
& 1.3 {$\times 10^{-11}$}
& 4.8 {$\times 10^{-11}$}
& 5.1 {$\times 10^{-8~}$}
& 3.9 {$\times 10^{-8~}$}
& 3.3 {$\times 10^{-11}$}
& 5.1 {$\times 10^{-10}$}
& 3.8 {$\times 10^{-11}$}
& ~8 {$\times 10^{-9~}$}
& 2.4 {$\times 10^{-11}$}
& 1.8 {$\times 10^{-9~}$}
\\
          0.05 -- 0.06 
& 3.6 {$\times 10^{-10}$}
& 2.2 {$\times 10^{-12}$}
& ~2 {$\times 10^{-9~}$}
& 4.2 {$\times 10^{-12}$}
& 8.5 {$\times 10^{-11}$}
& 1.6 {$\times 10^{-10}$}
& 1.1 {$\times 10^{-8~}$}
& 3.6 {$\times 10^{-11}$}
& 1.1 {$\times 10^{-9~}$}
& \underline{2.3 {$\times 10^{-13}$}}
& 7.8 {$\times 10^{-8~}$}
& 1.1 {$\times 10^{-11}$}
& 1.3 {$\times 10^{-12}$}
& 1.3 {$\times 10^{-7~}$}
\\
             0.06 -- 0.07 
& ~3 {$\times 10^{-9~}$}
& 1.1 {$\times 10^{-10}$}
& 5.5 {$\times 10^{-12}$}
& 8.1 {$\times 10^{-10}$}
& 4.7 {$\times 10^{-10}$}
& 4.2 {$\times 10^{-10}$}
& 5.4 {$\times 10^{-7~}$}
& 4.1 {$\times 10^{-7~}$}
& 4.3 {$\times 10^{-11}$}
& 1.2 {$\times 10^{-8~}$}
& 4.2 {$\times 10^{-11}$}
& 3.5 {$\times 10^{-9~}$}
& 1.6 {$\times 10^{-8~}$}
& 1.2 {$\times 10^{-8~}$}
\\
  0.07 -- 0.08
& 2.6 {$\times 10^{-13}$}
& ~8 {$\times 10^{-11}$}
& 1.4 {$\times 10^{-8~}$}
& 2.4 {$\times 10^{-10}$}
& 3.5 {$\times 10^{-10}$}
& 6.3 {$\times 10^{-9~}$}
& 4.8 {$\times 10^{-8~}$}
& 5.8 {$\times 10^{-10}$}
& 1.4 {$\times 10^{-8~}$}
& 4.3 {$\times 10^{-11}$}
& ~4 {$\times 10^{-9~}$}
& ~2 {$\times 10^{-10}$}
& 6.9 {$\times 10^{-9~}$}
& 3.3 {$\times 10^{-8~}$}
\\
                  0.08 -- 0.09
& 6.1 {$\times 10^{-8~}$}
& 1.3 {$\times 10^{-8~}$}
& 1.1 {$\times 10^{-9~}$}
& 2.3 {$\times 10^{-9~}$}
& 2.9 {$\times 10^{-12}$}
& 1.8 {$\times 10^{-11}$}
& 3.7 {$\times 10^{-7~}$}
& 5.5 {$\times 10^{-9~}$}
& 8.8 {$\times 10^{-11}$}
& 1.5 {$\times 10^{-9~}$}
& ~8 {$\times 10^{-10}$}
& 5.2 {$\times 10^{-10}$}
& 2.8 {$\times 10^{-12}$}
& 3.9 {$\times 10^{-7~}$}
\\
                  0.09 -- 0.1~~
& 1.3 {$\times 10^{-9~}$}
& 3.5 {$\times 10^{-7~}$}
& 2.8 {$\times 10^{-13}$}
& 3.1 {$\times 10^{-9~}$}
& 1.6 {$\times 10^{-9~}$}
& 4.5 {$\times 10^{-9~}$}
& 1.5 {$\times 10^{-7~}$}
& 1.2 {$\times 10^{-8~}$}
& 7.3 {$\times 10^{-8~}$}
& 2.4 {$\times 10^{-9~}$}
& 1.7 {$\times 10^{-11}$}
& 6.4 {$\times 10^{-9~}$}
& 3.6 {$\times 10^{-10}$}
& 4.4 {$\times 10^{-8~}$}
\\
      \bottomrule
   \end{tabular}
} 
\caption{Same as in Table~\ref{tab:pvals-sex} but on a set of narrower spectral bands spanning spectral range $[0, 0.1]$; results on spectral band 0 -- 0.1 replicated from Table~\ref{tab:pvals-sex}. Values lower than that obtained on corresponding tests at spectral band $[0,0.1]$ are underlined.} 
\label{tab:pvals-sex-lowerend}
\vspace{-3mm}
\end{table*}

\vspace{-2mm}
\section{Conclusions}
The design of cerebral cortical graphs, consisting of global hemisphere graphs and localized cortical graphs, was presented. Global hemisphere graphs encode the global topology cerebral hemisphere cortices, whereas local cortical graphs capture more subtle localized variations in cortical morphology. The set of spectra of local cortical graphs can be seen as an implicit decomposition of the single spectrum of the associated global hemisphere graph. Experimental results suggest the benefit of spectral features of cortical graphs as a powerful means for discriminative characterization of cortical morphology in relation to gender. Our future work will focus on testing the proposed cortical graph features on a larger cohort of healthy as well as patient subjects; in particular, characterization and early detection of changes in cortical morphology that arise in Alzheimer's disease \cite{Falahati2014} will be explored. The proposed cortical graphs can also be found applicable for graph spectral processing of functional MRI data, see for example \cite{Behjat2015}, in particular, through exploiting novel spectral graph filter design algorithms \cite{Behjat2016} that allow adaptation to both cortical structure as well as graph spectral content \cite{Stankovic2018} of cortical activity.


\vspace{-3mm}
\bibliographystyle{IEEEbib}
\bibliography{hbehjat_bibliography}

\begin{thebibliography}{10}

\bibitem{Decarli2005}
C.~DeCarli, J.~Massaro, D.~Harvey, J.~Hald, M.~Tullberg, R.~Au, A.~Beiser,
  R.~D'Agostino, and P.A. Wolf,
\newblock ``Measures of brain morphology and infarction in the framingham heart
  study: establishing what is normal,''
\newblock {\em Neurobiol. Aging}, vol. 26, no. 4, pp. 491--510, 2005.

\bibitem{Fischl2000}
B.~Fischl and A.M. Dale,
\newblock ``Measuring the thickness of the human cerebral cortex from magnetic
  resonance images,''
\newblock {\em Proc. Natl Acad. Sci.}, vol. 97, no. 20, pp. 11050--11055, 2000.

\bibitem{Gerardin2009}
E.~Gerardin, G.~Ch{\'e}telat, M.~Chupin, R.~Cuingnet, B.~Desgranges, H.~Kim,
  M.~Niethammer, B.~Dubois, S.~Leh{\'e}ricy, L.~Garnero, et~al.,
\newblock ``Multidimensional classification of hippocampal shape features
  discriminates {Alzheimer's} disease and mild cognitive impairment from normal
  aging,''
\newblock {\em Neuroimage}, vol. 47, no. 4, pp. 1476--1486, 2009.

\bibitem{Shen2012}
K.~Shen, J.~Fripp, F.~M{\'e}riaudeau, G.~Ch{\'e}telat, O.~Salvado, P.~Bourgeat,
  Alzheimer's Disease~Neuroimaging Initiative, et~al.,
\newblock ``Detecting global and local hippocampal shape changes in
  {Alzheimer's} disease using statistical shape models,''
\newblock {\em Neuroimage}, vol. 59, no. 3, pp. 2155--2166, 2012.

\bibitem{Wachinger2015}
C.~Wachinger, P.~Golland, W.~Kremen, B.~Fischl, M.~Reuter, Alzheimer's
  Disease~Neuroimaging Initiative, et~al.,
\newblock ``Brainprint: a discriminative characterization of brain
  morphology,''
\newblock {\em Neuroimage}, vol. 109, pp. 232--248, 2015.

\bibitem{Reuter2006}
M.~Reuter, F-E. Wolter, and N.~Peinecke,
\newblock ``{Laplace--Beltrami} spectra as {Shape-DNA} of surfaces and
  solids,''
\newblock {\em Comput. Aided Des}, vol. 38, no. 4, pp. 342--366, 2006.

\bibitem{Behjat2013}
H.~Behjat, N.~Leonardi, and D.~{Van De Ville},
\newblock ``Statistical parametric mapping of functional {MRI} data using
  wavelets adapted to the cerebral cortex,''
\newblock in {\em Proc. IEEE Int. Symp. Biomed. Imaging}, 2013, pp. 1070--1073.

\bibitem{Behjat2014}
H.~Behjat, N.~Leonardi, L.~S\"{o}rnmo, and D.~{Van De Ville},
\newblock ``Canonical cerebellar graph wavelets and their application to {fMRI}
  activation mapping,''
\newblock in {\em Proc. IEEE Int. Conf. Eng. Med. Biol. Soc.}, 2014, pp.
  1039--1042.

\bibitem{Behjat2015}
H.~Behjat, N.~Leonardi, L.~S\"{o}rnmo, and D.~{Van De Ville},
\newblock ``Anatomically-adapted graph wavelets for improved group-level {fMRI}
  activation mapping,''
\newblock {\em Neuroimage}, vol. 123, pp. 185--199, 2015.

\bibitem{Fischl2012}
B.~Fischl,
\newblock ``Freesurfer,''
\newblock {\em Neuroimage}, vol. 62, no. 2, pp. 774--781, 2012.

\bibitem{Luxburg2007}
U.~Von~Luxburg,
\newblock ``A tutorial on spectral clustering,''
\newblock {\em Stat. Comput.}, vol. 17, no. 4, pp. 395--416, 2007.

\bibitem{chacoUserguide}
Bruce Hendrickson and Robert Leland,
\newblock ``The chaco users guide. version 1.0,''
\newblock Tech. {R}ep., Sandia National Labs., Albuquerque, NM (United States),
  1993.

\bibitem{Parlett1998}
Beresford~N Parlett,
\newblock {\em The symmetric eigenvalue problem}, vol.~20,
\newblock Siam, 1998.

\bibitem{Shuman2015ieee}
D.~I. Shuman, C.~Wiesmeyr, N.~Holighaus, and P.~Vandergheynst,
\newblock ``Spectrum-adapted tight graph wavelet and vertex-frequency frames,''
\newblock {\em IEEE Trans. Signal Process.}, vol. 63, no. 16, pp. 4223--4235,
  2015.

\bibitem{HCP}
D.C. {Van Essen}, S.M. Smith, D.M. Barch, T.E. Behrens, E.~Yacoub, K.~Ugurbil,
  and WU-Minn~HCP Consortium.,
\newblock ``The {WU-Minn} human connectome project: an overview.,''
\newblock {\em Neuroimage}, vol. 80, pp. 62--79, 2013.

\bibitem{Kong2018}
Xiang-Zhen Kong, Samuel~R Mathias, Tulio Guadalupe, David~C Glahn, Barbara
  Franke, Fabrice Crivello, Nathalie Tzourio-Mazoyer, Simon~E Fisher, Paul~M
  Thompson, Clyde Francks, et~al.,
\newblock ``Mapping cortical brain asymmetry in 17,141 healthy individuals
  worldwide via the enigma consortium,''
\newblock {\em Proc. Natl Acad. Sci.}, vol. 115, no. 22, pp. E5154--E5163,
  2018.

\bibitem{Ritchie2018}
S.J. Ritchie, S.R. Cox, X.~Shen, M.V. Lombardo, L.M. Reus, C.~Alloza, M.A.
  Harris, H.L. Alderson, S.~Hunter, E.~Neilson, et~al.,
\newblock ``Sex differences in the adult human brain: evidence from 5216 {UK}
  {Biobank} participants,''
\newblock {\em Cerebral Cortex}, vol. 28, no. 8, pp. 2959--2975, 2018.

\bibitem{Falahati2014}
F.~Falahati, E.~Westman, and A.~Simmons,
\newblock ``Multivariate data analysis and machine learning in {Alzheimer's}
  disease with a focus on structural magnetic resonance imaging,''
\newblock {\em J. Alzheimers Dis.}, vol. 41, no. 3, pp. 685--708, 2014.

\bibitem{Behjat2016}
H.~Behjat, U.~Richter, D.~{Van De Ville}, and L.~S\"{o}rnmo,
\newblock ``Signal-adapted tight frames on graphs.,''
\newblock {\em IEEE Trans. Signal Process.}, vol. 64, no. 22, pp. 6017--6029,
  2016.

\bibitem{Stankovic2018}
L.~Stankovi{\'c}, E.~Sejdi{\'c}, and M.~Dakovi{\'c},
\newblock ``Reduced interference vertex-frequency distributions,''
\newblock {\em IEEE Signal Process. Lett.}, vol. 25, no. 9, pp. 1393--1397,
  2018.

\end{thebibliography}

\end{document}